\documentclass[a4paper]{article}

\usepackage{INTERSPEECH2019}
\usepackage[acronym,nomain,nonumberlist]{glossaries}
\usepackage{hyperref}
\newacronym{dl}{DL}{Deep Learning}
\newacronym{dbn}{DBN}{Deep Belief Network}
\newacronym{dnn}{DNN}{Deep Neural Network}
\newacronym{rbm}{RBM}{Restricted Boltzmann Machine}
\newacronym{cnn}{CNN}{Convolutional Neural Network}
\newacronym{rnn}{RNN}{Recurrent Neural Network}
\newacronym{lstm}{LSTM}{Long Short-Term Memory}
\newacronym{dbm}{DBM}{Deep Boltzmann Machine}
\newacronym{cd}{CD}{Contrastive Divergence}
\newacronym{urbm}{URBM}{Universal RBM}
\newacronym{udbn}{UDBN}{Universal DBN}
\newacronym{relu}{ReLU}{Rectified Linear Unit}
\newacronym{vrelu}{VReLU}{Variable ReLU}
\newacronym{ubm}{UBM}{Universal Background Model}
\newacronym{gmm}{GMM}{Gaussian Mixture Model}
\newacronym{map}{MAP}{Maximum a Posteriori}
\newacronym{nist}{NIST}{National Institute of Standard and Technology}
\newacronym{sre}{SRE}{Speaker Recognition Evaluation}
\newacronym{nap}{NAP}{Nuisance Attribute Projection}
\newacronym{jfa}{JFA}{Joint Factor Analysis}
\newacronym{lid}{LID}{Language Identification}
\newacronym{lda}{LDA}{Linear Discriminant Analysis}
\newacronym{vad}{VAD}{Voice Activity Detection}
\newacronym{plda}{PLDA}{Probabilistic Linear Discriminant Analysis}
\newacronym{wccn}{WCCN}{Within-Class Covariance Normalization}
\newacronym{asr}{ASR}{Automatic Speech Recognition}
\newacronym{det}{DET}{Detection Error Tradeoff}
\newacronym{eer}{EER}{Equal Error Rate}
\newacronym{dcf}{DCF}{Detection Cost Function}
\newacronym{mindcf}{minDCF}{minimum DCF}
\newacronym{far}{FAR}{False Acceptance Rate}
\newacronym{frr}{FRR}{False Rejection Rate}
\newacronym{fbe}{FBE}{Filter Bank Energy}
\newacronym{ff}{FF}{Frequency Filtering}
\newacronym{hmm}{HMM}{Hidden Markov Model}
\newacronym{lpc}{LPC}{Linear Predictive Coefficient}
\newacronym{lpr}{LPR}{Log Posterior Ratio}
\newacronym{mfcc}{MFCC}{Mel-Frequency Cepstral Coefficient}
\newacronym{lfcc}{LFCC}{Linear Frequency Cepstral Coefficient}
\newacronym{nn}{NN}{Neural Network}
\newacronym{pca}{PCA}{Principal Component Analysis}
\newacronym{plp}{PLP}{Perceptual Linear Predictive}
\newacronym{sdc}{SDC}{Shifted Delta Cepstrum}
\newacronym{psd}{PSD}{Power Spectral Density}
\newacronym{rbf}{RBF}{Radial Basis Function}
\newacronym{dct}{DCT}{Discrete Cosine Transform}
\newacronym{snr}{SNR}{Signal-to-Noise Ratio}
\newacronym{svm}{SVM}{Support Vector Machine}
\newacronym{cmn}{CMN}{Cepstral Mean Normalization}
\newacronym{cmvn}{CMVN}{Cepstral Mean and Variance Normalization}
\newacronym{cdf}{CDF}{Cumulative Distribution Function}
\newacronym{pdf}{PDF}{Probability Density Function}
\newacronym{em}{EM}{Expectation-Maximization}
\newacronym{sv}{SV}{Speaker Verification}
\newacronym{sr}{SR}{Speaker Recognition}
\newacronym{an}{AN}{Adversarial Network}
\newacronym{en}{EN}{Encoding Network}
\newacronym{dn}{DN}{Discriminative Network}
\newacronym{mlp}{MLP}{Multilayer Perceptron}
\newacronym{gru}{GRU}{Gated Recurrent Unit}
\newacronym{tdnn}{TDNN}{Time Delay Neural Network}
\newacronym{nmt}{NMT}{Neural Machine Translation}
\newacronym{fc}{FC}{Fully Connected}

\title{Self Multi-Head Attention for Speaker Recognition}
\name{Miquel India$^{\dagger}$, Pooyan Safari$^{\dagger}$, Javier Hernando}
\address{Universitat Politecnica de Catalunya, Barcelona, Spain}
\email{$^{\dagger}$first.last@tsc.upc.edu, javier.hernando@upc.edu}

\begin{document}

\maketitle

\begin{abstract}

Most state-of-the-art \gls{dl} approaches for speaker recognition work on a short utterance level. Given the speech signal, these algorithms extract a sequence of speaker embeddings from short segments and those are averaged to obtain an utterance level speaker representation. In this work we propose the use of an attention mechanism to obtain a discriminative speaker embedding given non fixed length speech utterances. Our system is based on a \gls{cnn} that encodes short-term speaker features from the spectrogram and a self multi-head attention model that maps these representations into a long-term speaker embedding. The attention model that we propose produces multiple alignments from different subsegments of the \gls{cnn} encoded states over the sequence. Hence this mechanism works as a pooling layer which decides the most discriminative features over the sequence to obtain an utterance level representation. We have tested this approach for the verification task for the VoxCeleb1 dataset. The results show that self multi-head attention outperforms both temporal and statistical pooling methods with a $18\%$ of relative EER. Obtained results show a $58\%$ relative improvement in EER compared to i-vector+PLDA.

\end{abstract}


\section{Introduction}


Recently there have been several attempts to apply Deep Learning (DL) in order to build speaker embeddings. Speaker embedding is often referred to a single low dimensional vector representation of the speaker characteristics from a speech signal extracted using a \gls{nn} model. For text-independent \gls{sr}, which is the focus of this work, these models can be trained either in a supervised (e.g., \cite{variani2014deep,snyder2017deep}) or in an unsupervised (e.g., \cite{vasilakakis2013speaker,safari2016features}) fashion. Supervised speaker embeddings are produced by training a deep architecture using speaker-labeled background data. This network, which is capable to produce high-level features, is usually trained to discriminate the background speakers. Then in the testing phase, the output layer is discarded, the feature vectors of an unknown speaker are given through the network, and the pooled representation from the activation of a given hidden layer are considered as the speaker embedding \cite{variani2014deep,snyder2017deep,bhattacharya2017deep,snyder2018x}. The reported results from different works have shown that, in most of cases, the largest improvements are obtained on short utterances compared to the conventional i-vectors \cite{bhattacharya2017deep,snyder2017deep}. That suggests
that \gls{dl} technology can model the speaker characteristics of a short-duration speech signal better than the traditional signal processing techniques.

In \cite{variani2014deep}, the inputs of the network are the speaker feature vectors stacked over a context window. They use a \gls{dnn} with max-pooling and dropout regularization applied on the last two hidden layers. There are also other works which employes other deep architectures such as \gls{cnn} and \gls{tdnn} \cite{bhattacharya2017deep,li2018full}. Snyder \textit{et al.}, in \cite{snyder2016deep} introduced a temporal pooling to extract speaker embeddings and a \gls{dnn} architecture with a PLDA-like objective function. This function operates on pairs of embeddings to maximize the probability for the embeddings of the same speakers and minimize it otherwise. In \cite{snyder2017deep}, they take advantage of a \gls{tdnn} which is further followed by a statistical pooling and a \gls{dnn} classifier. The statistical pooling layer aggregates input segments over the variable-length and prepares the fixed-dimensional statistics vectors as the inputs of a feed-forward network. The second part of the network has only two hidden layers whose activations can be used as speaker embeddings. The preliminary results showed that these embeddings outperform the traditional i-vectors \cite{dehak2011front} for short duration speech segments \cite{snyder2017deep}. However, \cite{snyder2018x} a recent work has shown that data augmentation, consisting of added noise and reverberation, can significantly improve the performance of these embeddings (\textit{x-vectors} as they referred to), while it is not so effective for i-vectors \cite{snyder2018x}. There have also been some efforts to improve the quality and generalization powers of \textit{x-vectors} by the modification applied to the network architecture \cite{novoselov2018deep} and the training procedure \cite{li2018gaussian,zeinali2018improve,huang2018angular}.



Attention mechanisms are one of the main reasons of the success of sequence-to-sequence (seq2seq) models in tasks like \gls{nmt} or \gls{asr} \cite{bahdanau2014neural,vaswani2017attention,chan2015listen}. In seq2seq models, these algorithms are applied over the encoded sequence in order to help the decoder to decide which region of the sequence must be either translated or recognized. For speaker recognition, these models have been also used for text-dependent speaker verification. In works like \cite{bhattacharya2017deep, zhang2016end, chowdhury2017attention}, attention is applied over the hidden states of a \gls{rnn} in order to pool these states into speaker embeddings. The same idea has been also used for text-independent verification.  In \cite{cai2018exploring}, a unified framework is introduced for both speaker and language recognition. In this architecture, variable-length input utterance is fed into a network that encodes an utterance level representation. In addition to temporal pooling, they have also adopted a self-attention pooling mechanism and a learnable dictionary encoding layer to get the utterance level representation. Multi-head attention is a newly emerging attention mechanism which is originally proposed in \cite{vaswani2017attention} for a \textit{Transformer} architecture and appeared very effective in many seq2seq models such as \cite{vaswani2017attention,chiu2018state,dehghani2018universal}.


In this paper we present a multi-head attention based network for speaker verification. This mechanism is used as a self attentive pooling layer to create an utterance level embedding. Given a set of encoded representations from a \gls{cnn} feature extractor, self attention performs a weighted average of these representations. This mechanism differs from other pooling methods in that the average weights are also trained as network parameters. In comparison with other works like \cite{cai2018exploring}, our approach introduces a major improvement by using multi-head attentions (instead of single-head attention as in \cite{cai2018exploring}). This allows the model to attend to different parts of the sequence, which is one of the main limitations of vanilla self attentive pooling. In the same way, multi-head also helps the network to attend to different sub-sets of the encoded representations. Therefore the encoder is not forced to create overall speaker embeddings from the feature level. Attention allows the encoder to create different sets of features, so the model can attend to the most discriminative patterns from different positions of the sequence.
The main contribution of this works is the introduction of a pooling layer which takes advantage of multi-head self attention mechanism to create more discriminative speaker embeddings. We compare this pooling mechanism with  temporal and statistical pooling layers. In order to show the effectiveness of the proposed approach, these embeddings will be assessed in a text-independent speaker verification task.    


The rest of this paper is structured as follows. Section 2 explains self multi-head attention pooling. Section 3 illustrates the architecture of the system. Section 4 gives the details of the system setup. Experimental results are presented in section 5. The concluding remarks and some future works are given in section 6.

\section{Self Multi-Head Attention Pooling}

Self attentive pooling attention was initially proposed in \cite{cai2018exploring} for text-independent speaker verification. Their objective was to use a trainable and more adapted layer for pooling than vanilla temporal average. Given a sequence of encoded hidden states from a network, temporal pooling averages these representations over the time to obtain a final encoded embedding. The main problem of this method is that assumes that all the elements of the sequence must contribute equally in obtaining the utterance level representation. Self attentive pooling is a mechanism that through a trainable layer is able to assign a weight over each representation of the sequence. Hence given these weights, the utterance level representation is obtained through the respective weighted average of these representations.

Consider a sequence of hidden of sequence states $h=[h_1 h_2  ... h_N]$, with $h_t\in \mathbb{R}^{d} $, and a trainable $u \in \mathbb{R}^d$. We can define a relevance scalar weight for each element of the sequence trough a softmax layer:

\begin{equation}
\centering
w_t = \frac{\exp{(h^{T}_{t}u)}}{\sum^{N}_{l=1}\exp{(h^{T}_{l}u)}}
\end{equation}

Given the set of weights over all the elements of the sequence, we can then obtain the pooled representation as the  weighted average of the hidden states:

\begin{equation}
\centering
c = \sum^{N}_{t=1}h^{T}_{t}w_t
\end{equation}

This attention mechanism has some limitations. The main restriction is that attention weights are calculated considering the whole information of the embedding. Therefore, we assume that all the discriminative information of the signal must come from the same encoded representations of the utterance.

\begin{figure}[!t]
\centering
\includegraphics[width=3.2in]{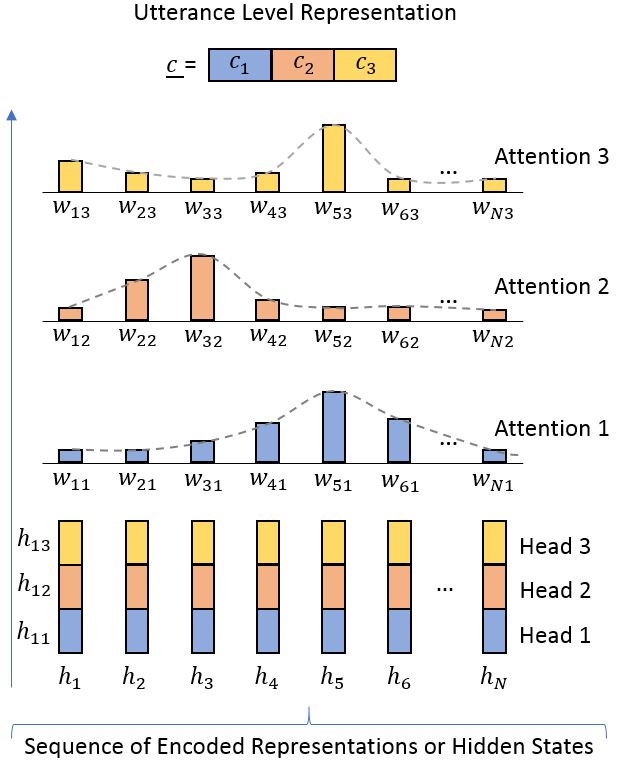}
\caption{An example of the Self Multi-Head Attention Pooling with $3$ heads.}
\label{FIG: multihead-attn}
\end{figure}

Multi-head attention model was firstly introduced in \cite{vaswani2017attention}. This approach consists on splitting the encoded representations of the sequence into homogeneous sub-vectors called heads (Figure \ref{FIG: multihead-attn}). If we consider a number of $k$ heads for the multi-head attention, now $h_t=[h_{t1}h_{t2}...h_{tk}]$ where $h_{tj} \in \mathbb{R}^{d/k}$. We can compute then the head size as $d/k$. In the same way we have also a trainable parameter $u=[u_1u_2...u_k]$ where $u_j  \in \mathbb{R}^{d/k}$. A different attention is then applied over each head of the encoded sequence:

\begin{equation}
\centering
w_{tj} = \frac{\exp{(h^{T}_{tj}u_j)}}{\sum^{N}_{l=1}\exp{(h^{T}_{lj}u_j)}}
\end{equation}

where $w_{tj}$ corresponds to the attention weight of the head $j$ on the step $t$ of the sequence. If each head corresponds to a subspace of the hidden state, the weight sequence of that head can be considered as a probability distribution function from that sub-space features over the sequence. We then compute a new pooled representation for each head in the same way than vanilla self attention:

\begin{equation}
\centering
c_j = \sum^{N}_{t=1}h^{T}_{tj}w_{tj}
\end{equation}

where $c_j\in \mathbb{R}^{d/k}$ corresponds to the utterance level representation from head $j$. The final utterance level representation is then obtained with the concatenation of the utterance level vectors from all the heads $c=[c_1c_2...c_k]$. This method allows the network to extract different kind information over different regions of the network. Besides, the main advantage of this attention variation is that it does not increase the complexity of the model adding more parameters on the model. Instead of having a global $u$ attention vector, we have now a subset of attention vector $u_j$ which sums the same number of components than $u$.




\section{System Description}

\begin{figure}[!t]
\centering
\includegraphics[width=3.3in]{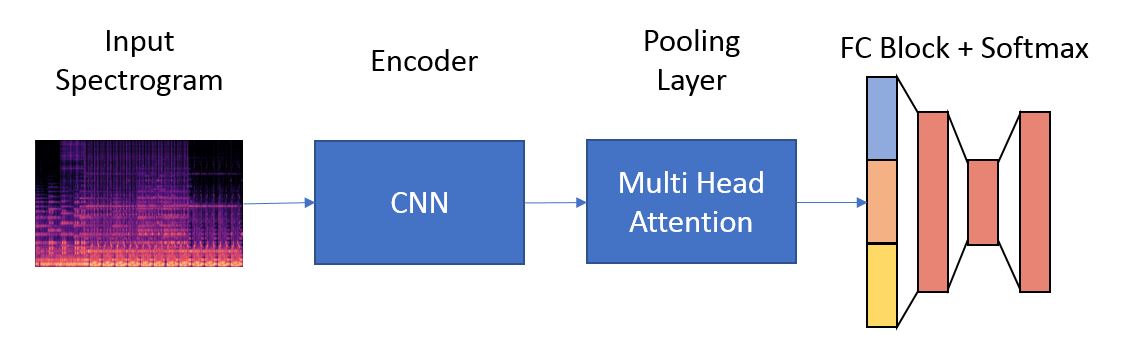}
\caption{System Diagram.}
\label{FIG: System Diagram}
\end{figure}

Figure \ref{FIG: System Diagram} shows the overall architecture used for this work. The proposed neural network is a \gls{cnn} based encoder and an attention based pooling layer followed by a set of dense layers. The network is fed with variable length mel-spectrogram features. These features are then mapped into a sequence of speaker representations trough a \gls{cnn} encoder. This \gls{cnn} feature extractor is based on  of the VGG proposed in \cite{hori2017advances} for the \gls{asr} task. In our case, we have extended this architecture so as to work for speaker verification. Our adapted VGG is composed of three convolution blocks, where each block contains two concatenated convolutional layers followed by a max pooling layer with a $2$x$2$ stride. Hence given a spectrogram a of $N$ frames, the VGG performs a down-sampling reducing  its output into a sequence of $N/8$ representations. Given this set of representations, the attention mechanism is then used to transform the encoded states of the \gls{cnn} into an overall speaker representation. Finally this fixed length embedding is feed into a set of \gls{fc} layers and a softmax layer. We refer to bottle neck layer previous to the softmax layer as the speaker embedding. The softmax layer corresponds to the speaker labels of the train partition corpus. Hence the network is trained as a speaker classifier. The speaker embedding layer corresponds to the speaker representation that will be used for the speaker verification task. 

\begin{table}[t]
\begin{center}
\begin{tabular}{|c|c|c|c|c|c|}
\hline
Layer & Size & In Dim. & Out Dim. & Stride & Feat Size\\ \hline
conv11 & 3x3 & 1 & 128 & 1x1 & 128xN\\ \hline
conv12 & 3x3 & 128 & 128 & 1x1 & 128xN\\ \hline
mpool1 & 2x2 & - & - & 2x2 & 64xN/2\\ \hline
conv21 & 3x3 & 128 & 256 & 1x1 & 64xN/2\\ \hline
conv22 & 3x3 & 256 & 256 & 1x1 & 64xN/2\\ \hline
mpool2 & 2x2 & - & - & 2x2 & 32xN/4\\ \hline
conv31 & 3x3 & 256 & 512 & 1x1 & 32xN/4\\ \hline
conv32 & 3x3 & 512 & 512 & 1x1 & 32xN/4\\ \hline
mpool3 & 2x2 & - & - & 2x2 & 16XN/8\\ \hline
 flatten & - & 512 & 1 & - & 8192xN/8\\ \hline
\end{tabular}
\end{center}
\caption{\gls{cnn} Architecture. In and Out Dim. refers to the input and output feature maps of the layer. Feat Size refers to the dimension of each one of this output feature maps.}
\label{TAB: Network Dimensions}
\end{table}

\section{Experimental Setup}

The proposed system in this work will be tested on VoxCeleb1 \cite{Nagrani17}. This corpus is a large multimedia database that contains over $100,00$ utterances for $1,251$ celebrities, extracted from videos uploaded to \textit{Youtube}. For each person in the corpus there is an average of $18$ videos. Each of these videos has been split into approximately $123$ short speech utterances of $8.2$ seconds average length. The proposed approaches will be evaluated on the original VoxCeleb1 speaker verification protocol \cite{Chung18b}. Hence the network is trained with VoxCeleb1 development partition and evaluated on the test set. 

Three different baselines will be considered to compare with the presented approach. The soft multi-head attention pooling will be evaluated against two statistical based methods: temporal and statistical pooling. In order to evaluate them, this pooling layers will replace the attention pooling block without modifying any other parameter of the network. The speaker vectors used for the verification tests will be extracted from the same speaker embedding layer for each of the pooling methods. The metric used to compute the scores between embeddings for the verification task is cosine distance.  Additionally we have also considered an i-vector + PLDA baseline \cite{dehak2011front, prince2007probabilistic}. The i-vector is created from $20$ MFCC + delta coefficients features. The extraction is performed using a $1024$ \gls{ubm} and a $400$ total variability matrix. G-PLDA\cite{prince2007probabilistic} is applied with $200$ eigenvector size. 

\begin{table}[!t]
    \caption{Evaluation results of the text-independent verification task on VoxCeleb 1. The results for our proposed architecture have been obtained using cosine scoring.}
   \label{TAB: Results}
    \centering
    \begin{tabular}{lcc}
        \toprule
        \textbf{Approach} & \textbf{DCF} &  \textbf{EER} \\ \hline
        I-vector + PLDA & 0.0078 & $9.54$\\ \hline
         CNN + Temporal Pooling & 0.0047 & $4.91$\\
         CNN + Statistical Pooling & 0.0046 & $4.9$\\ 
         CNN + Att. Pooling  & 0.005 & $4.71$\\
         CNN + MHA Pooling    &  \textbf{0.0045} & \textbf{4.0}\\ \hline
         \bottomrule 
    \end{tabular}
\end{table}

\begin{figure}[t]
\centering
\includegraphics[width=2.6in]{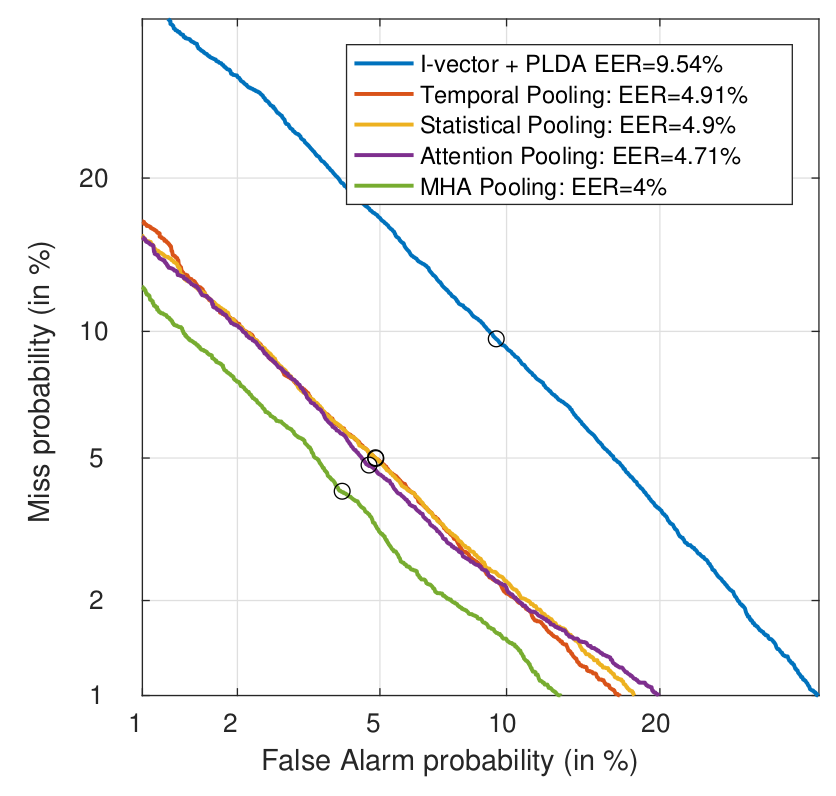}
\caption{DET curves for the experiments on VoxCeleb 1 verification task. MHA stands for the Multi-Head Attention.}
\label{FIG: DET-curves}
\end{figure}

\begin{figure*}[t]
\centering
\includegraphics[width=6.5in]{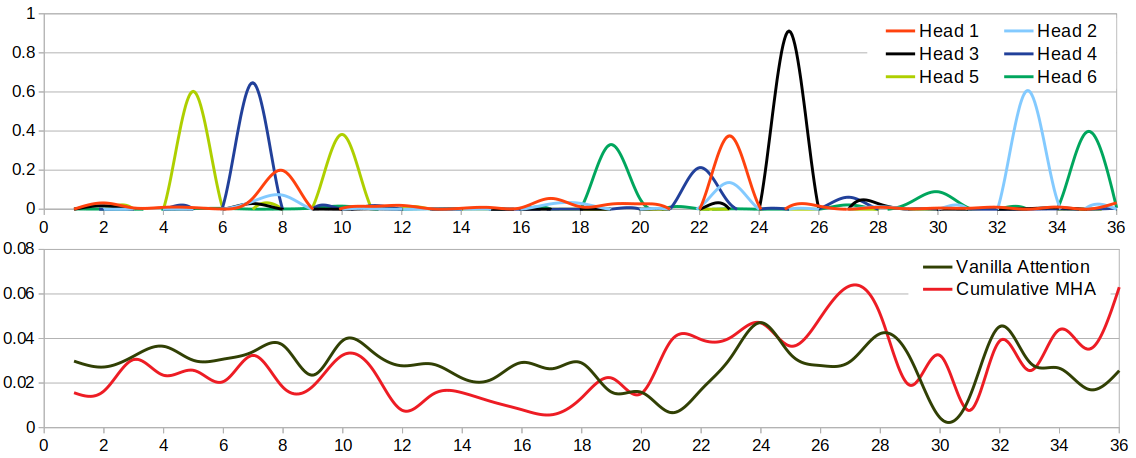}
\caption{\textbf{Top}: Analysis of the weight values for the first six multi-head attentions over a test utterance. \textbf{Bottom}: Comparison between vanilla attention weights and the averaged weights over all the heads of our proposed model (Cumulative MHA). The weights are extracted from the same test utterance than the top image.}
\label{FIG: Attention}
\end{figure*}

The proposed network has been trained to classify variable length speaker utterances. For feature extraction we have used librosa \cite{brian_mcfee_2019_2564164} to extract $128$ dimension mel-spectrograms. The \gls{cnn} encoder is then feed with $128$x$N$ spectrograms to obtain a sequence of $8192$x$N/8$ encoded hidden representations. The setup of the \gls{cnn} feature extractor can be found on Table \ref{TAB: Network Dimensions}. The pooling layer maps the encoded sequence into an unique speaker representation. The following \gls{fc} Block consists on two consecutive dense layers with $1024$ and $500$ dimension, where the last layer correspond to the speaker embedding. A final softmax layer is then fed with the speaker embedding. Batch normalization has been applied on the $1024$ dense layer and $0.2$ dropout on the softmax layer. Adam optimizer has been used to train all the models with standard values and learning rate of $1e-4$. Finally we have applied a $5$ epochs patience early stopping criterion.

\section{Results}

The proposed attention pooling layer has been evaluated with different approaches in the VoxCeleb1 text-independent verification task and presented in Table \ref{TAB: Results}. Performance is evaluated using the \gls{eer} and the minimum Decision Cost Function (DCF) calculated using $C_{FA}=1$, $C_{M}=1$, and $P_{T}=0.01 $. MHA Pooling refers to the best self multi-head attention model that we have trained. This model has a $128$ head size, which corresponds to $64$ heads per encoded representation. I-vector with PLDA have shown the worst results for this task. As it has mentioned before, i-vectors performance decreases in short-utterance condition. Following the i-vector, the statistical based pooling layers have scored $4.91\%$ and $4.9\%$ EER, respectively. Vanilla self attentive pooling performance has shown a $4.71\%$ EER. Similar to the work proposed in \cite{cai2018exploring}, self attentive pooling doesn't lead to a big improvement. Here it has only shown a $4\%$ improvement  relative improvement. Self Multi-Head attention has shown the best result of all the evaluated approaches. It outperforms both i-vector+PLDA and statistical based pooling layers with a $58\%$ EER and $18\%$ EER improvement, respectively. In comparison with self attentive pooling layer, MHA also shows a noticeable improvement of $14 \%$ EER. 

Figure \ref{FIG: DET-curves} shows the \gls{det} curves of the i-vector+PLDA baseline and our proposed architecture with different pooling mechanisms. It shows that not only at the EER but also at all other working points MHA pooling outperforms other pooling mechanisms by a large margin. 

In order to understand the improvement achieved with self-multihead attention pooling in comparison with vanilla attention models, we have assessed their attention weights. Figure \ref{FIG: Attention} shows the weight values created over the encoded features of the CNN for both vanilla and multi-head attention pooling in one of the VoxCeleb1 test utterances. On the top we can appreciate how each of the several heads of the multi-head model attends to  different regions of the sequence. That suggests that the model is able to capture sub-sets of features from the encoded representations in different parts of the signal. 
On the bottom, the weight values from vanilla attention model are compared with the averaged weights of the different heads of the multi-head model (cumulative multi-head). Vanilla attention weights have a more uniform distribution over the sequence than the weights showed by the heads of the multi-head model in the top image. If we compare the weight  alignment created from both vanilla attention and cumulative multi-head, several discriminative regions are commonly detected. However, there are some regions of the sequence attended by the MHA model that vanilla attention has not detected. That suggests that MHA degrees of freedom permits the detection procedure to focus on more specific regions of the sequence.

\section{Conclusions}

In this paper we have applied a self multi-head attention mechanism to obtain speaker embeddings at level utterance by pooling short-term features. This pooling layer have been tested in a neural network based on a CNN that maps spectrograms into sequences of speaker vectors. These vectors are then input to the pooling layer, which output activation is then connected to a set of dense layers. The network is trained as a speaker classifier and a bottleneck layer from the fully connected block is used as speaker embedding. We have evaluated this approach with other pooling methods for the text-independent verification task using the speaker embeddings and applying cosine distance. The presented approach have outperformed standard pooling methods based on statistical layers and vanilla attention models. We have also analyzed the multi-head attention alignments over a sequence. This analysis have shown that self multi head attention layer allows to capture specific sub-sets of features over different regions of a sequence.

\section{Acknowledgements}
This work was supported in part by the Spanish Project DeepVoice (TEC2015-69266-P).
\bibliographystyle{IEEEtran}

\bibliography{article}

\begin{thebibliography}{10}
\providecommand{\url}[1]{#1}
\csname url@samestyle\endcsname
\providecommand{\newblock}{\relax}
\providecommand{\bibinfo}[2]{#2}
\providecommand{\BIBentrySTDinterwordspacing}{\spaceskip=0pt\relax}
\providecommand{\BIBentryALTinterwordstretchfactor}{4}
\providecommand{\BIBentryALTinterwordspacing}{\spaceskip=\fontdimen2\font plus
\BIBentryALTinterwordstretchfactor\fontdimen3\font minus
  \fontdimen4\font\relax}
\providecommand{\BIBforeignlanguage}[2]{{%
\expandafter\ifx\csname l@#1\endcsname\relax
\typeout{** WARNING: IEEEtran.bst: No hyphenation pattern has been}%
\typeout{** loaded for the language `#1'. Using the pattern for}%
\typeout{** the default language instead.}%
\else
\language=\csname l@#1\endcsname
\fi
#2}}
\providecommand{\BIBdecl}{\relax}
\BIBdecl

\bibitem{variani2014deep}
E.~Variani, X.~Lei, E.~McDermott, I.~L. Moreno, and J.~Gonzalez-Dominguez,
  ``Deep neural networks for small footprint text-dependent speaker
  verification,'' in \emph{2014 IEEE International Conference on Acoustics,
  Speech and Signal Processing (ICASSP)}.\hskip 1em plus 0.5em minus
  0.4em\relax IEEE, 2014, pp. 4052--4056.

\bibitem{snyder2017deep}
D.~Snyder, D.~Garcia-Romero, D.~Povey, and S.~Khudanpur, ``Deep neural network
  embeddings for text-independent speaker verification,'' in
  \emph{Interspeech}, 2017, pp. 999--1003.

\bibitem{vasilakakis2013speaker}
V.~Vasilakakis, S.~Cumani, P.~Laface, and P.~Torino, ``Speaker recognition by
  means of deep belief networks,'' \emph{Proc. Biometric Technologies in
  Forensic Science}, 2013.

\bibitem{safari2016features}
P.~Safari, O.~Ghahabi, and F.~J. Hernando~Peric{\'a}s, ``From features to
  speaker vectors by means of restricted boltzmann machine adaptation,'' in
  \emph{ODYSSEY 2016-The Speaker and Language Recognition Workshop}, 2016, pp.
  366--371.

\bibitem{bhattacharya2017deep}
G.~Bhattacharya, M.~J. Alam, and P.~Kenny, ``Deep speaker embeddings for
  short-duration speaker verification,'' in \emph{Interspeech}, 2017, pp.
  1517--1521.

\bibitem{snyder2018x}
D.~Snyder, D.~Garcia-Romero, G.~Sell, D.~Povey, and S.~Khudanpur, ``X-vectors:
  Robust dnn embeddings for speaker recognition,'' in \emph{2018 IEEE
  International Conference on Acoustics, Speech and Signal Processing
  (ICASSP)}.\hskip 1em plus 0.5em minus 0.4em\relax IEEE, 2018, pp. 5329--5333.

\bibitem{li2018full}
L.~Li, Z.~Tang, D.~Wang, and T.~F. Zheng, ``Full-info training for deep speaker
  feature learning,'' in \emph{2018 IEEE International Conference on Acoustics,
  Speech and Signal Processing (ICASSP)}.\hskip 1em plus 0.5em minus
  0.4em\relax IEEE, 2018, pp. 5369--5373.

\bibitem{snyder2016deep}
D.~Snyder, P.~Ghahremani, D.~Povey, D.~Garcia-Romero, Y.~Carmiel, and
  S.~Khudanpur, ``Deep neural network-based speaker embeddings for end-to-end
  speaker verification,'' in \emph{2016 IEEE Spoken Language Technology
  Workshop (SLT)}.\hskip 1em plus 0.5em minus 0.4em\relax IEEE, 2016, pp.
  165--170.

\bibitem{dehak2011front}
N.~Dehak, P.~J. Kenny, R.~Dehak, P.~Dumouchel, and P.~Ouellet, ``Front-end
  factor analysis for speaker verification,'' \emph{IEEE Transactions on Audio,
  Speech, and Language Processing}, vol.~19, no.~4, pp. 788--798, 2011.

\bibitem{novoselov2018deep}
S.~Novoselov, A.~Shulipa, I.~Kremnev, A.~Kozlov, and V.~Shchemelinin, ``On deep
  speaker embeddings for text-independent speaker recognition,'' \emph{arXiv
  preprint arXiv:1804.10080}, 2018.

\bibitem{li2018gaussian}
L.~Li, Z.~Tang, Y.~Shi, and D.~Wang, ``Gaussian-constrained training for
  speaker verification,'' \emph{arXiv preprint arXiv:1811.03258}, 2018.

\bibitem{zeinali2018improve}
H.~Zeinali, L.~Burget, J.~Rohdin, T.~Stafylakis, and J.~Cernocky, ``How to
  improve your speaker embeddings extractor in generic toolkits,'' \emph{arXiv
  preprint arXiv:1811.02066}, 2018.

\bibitem{huang2018angular}
Z.~Huang, S.~Wang, and K.~Yu, ``Angular softmax for short-duration
  text-independent speaker verification,'' \emph{Proc. Interspeech, Hyderabad},
  2018.

\bibitem{bahdanau2014neural}
D.~Bahdanau, K.~Cho, and Y.~Bengio, ``Neural machine translation by jointly
  learning to align and translate,'' \emph{arXiv preprint arXiv:1409.0473},
  2014.

\bibitem{vaswani2017attention}
A.~Vaswani, N.~Shazeer, N.~Parmar, J.~Uszkoreit, L.~Jones, A.~N. Gomez,
  {\L}.~Kaiser, and I.~Polosukhin, ``Attention is all you need,'' in
  \emph{Advances in Neural Information Processing Systems}, 2017, pp.
  5998--6008.

\bibitem{chan2015listen}
W.~Chan, N.~Jaitly, Q.~V. Le, and O.~Vinyals, ``Listen, attend and spell,''
  \emph{arXiv preprint arXiv:1508.01211}, 2015.

\bibitem{zhang2016end}
S.-X. Zhang, Z.~Chen, Y.~Zhao, J.~Li, and Y.~Gong, ``End-to-end attention based
  text-dependent speaker verification,'' in \emph{2016 IEEE Spoken Language
  Technology Workshop (SLT)}.\hskip 1em plus 0.5em minus 0.4em\relax IEEE,
  2016, pp. 171--178.

\bibitem{chowdhury2017attention}
F.~Chowdhury, Q.~Wang, I.~L. Moreno, and L.~Wan, ``Attention-based models for
  text-dependent speaker verification,'' \emph{arXiv preprint
  arXiv:1710.10470}, 2017.

\bibitem{cai2018exploring}
W.~Cai, J.~Chen, and M.~Li, ``Exploring the encoding layer and loss function in
  end-to-end speaker and language recognition system,'' in \emph{Proc. Odyssey
  2018 The Speaker and Language Recognition Workshop}, 2018, pp. 74--81.

\bibitem{chiu2018state}
C.-C. Chiu, T.~N. Sainath, Y.~Wu, R.~Prabhavalkar, P.~Nguyen, Z.~Chen,
  A.~Kannan, R.~J. Weiss, K.~Rao, E.~Gonina \emph{et~al.}, ``State-of-the-art
  speech recognition with sequence-to-sequence models,'' in \emph{2018 IEEE
  International Conference on Acoustics, Speech and Signal Processing
  (ICASSP)}.\hskip 1em plus 0.5em minus 0.4em\relax IEEE, 2018, pp. 4774--4778.

\bibitem{dehghani2018universal}
M.~Dehghani, S.~Gouws, O.~Vinyals, J.~Uszkoreit, and {\L}.~Kaiser, ``Universal
  transformers,'' \emph{arXiv preprint arXiv:1807.03819}, 2018.

\bibitem{hori2017advances}
T.~Hori, S.~Watanabe, Y.~Zhang, and W.~Chan, ``Advances in joint ctc-attention
  based end-to-end speech recognition with a deep cnn encoder and rnn-lm,''
  \emph{arXiv preprint arXiv:1706.02737}, 2017.

\bibitem{Nagrani17}
A.~Nagrani, J.~S. Chung, and A.~Zisserman, ``Voxceleb: a large-scale speaker
  identification dataset,'' in \emph{INTERSPEECH}, 2017.

\bibitem{Chung18b}
J.~S. Chung, A.~Nagrani, and A.~Zisserman, ``Voxceleb2: Deep speaker
  recognition,'' in \emph{INTERSPEECH}, 2018.

\bibitem{prince2007probabilistic}
S.~J. Prince and J.~H. Elder, ``Probabilistic linear discriminant analysis for
  inferences about identity,'' in \emph{2007 IEEE 11th International Conference
  on Computer Vision}.\hskip 1em plus 0.5em minus 0.4em\relax IEEE, 2007, pp.
  1--8.

\bibitem{brian_mcfee_2019_2564164}
\BIBentryALTinterwordspacing
B.~McFee, M.~McVicar, S.~Balke, V.~Lostanlen, C.~Thomé, C.~Raffel, D.~Lee,
  K.~Lee, O.~Nieto, F.~Zalkow, D.~Ellis, E.~Battenberg, R.~Yamamoto, J.~Moore,
  Z.~Wei, R.~Bittner, K.~Choi, nullmightybofo, P.~Friesch, F.-R. Stöter,
  Thassilo, M.~Vollrath, S.~K. Golu, nehz, S.~Waloschek, Seth, R.~Naktinis,
  D.~Repetto, C.~F. Hawthorne, and C.~Carr, ``librosa/librosa: 0.6.3,'' Feb.
  2019. [Online]. Available: \url{https://doi.org/10.5281/zenodo.2564164}
\BIBentrySTDinterwordspacing

\end{thebibliography}

\end{document}